\documentclass{PoS}

\title{\center{Lecture about the Recent Nobel Prize} \\
       \normalsize --- From B Factory to the Large Hadron Collider ---}

\ShortTitle{Nobel\,\ 2008}

\author{\speaker{George Wei-Shu Hou}\thanks{
 I thank the organizers for the invitation.
 This work is supported in part by
 the National Science Council of R.O.C.
 grant NSC 97-2112-M-002-004-MY3,
 by National Taiwan University
 grant 97R0066-60, and
 the National Center of Theoretical Sciences (North).
 }\\
        National Taiwan University\\
        E-mail: \email{wshou@phys.ntu.edu.tw}}

\abstract{These are the transcriptions of ``a talk describing the
theoretical insights of those honored (and the one who wasn't) and
how this has led to all the physics we have been doing over the
last couple of decades.''
After prologue, we first deal with Nambu's insight on Spontaneous
Symmetry Breaking, which is rooted in an analogy with the BCS
theory of superconductivity. The insight resonates to this day, as
we await the LHC era to dawn.
The second half starts from Gell-Mann--L\'evy--Cabibbo theory,
through the GIM mechanism that completed the $2\times 2$
rotations, to the insight of Kobayashi and Maskawa that CP
violation could arise from the charged currents, if there exists a
3rd generation of quarks. The richness that followed defines this
(FPCP) conference.
We end with a perspective on a (possible) redux with a 4th
generation of quarks. }

\FullConference{Flavor Physics and CP Violation 2009,\\
        May 27 - June 1, 2009,\\
        Lake Placid, NY, USA}

\begin{document}

\noindent{\bf 0. \ \ Prologue: \ Predicting the 2008 Prize}
 \vskip0.2cm

On September 30th, 2008, Ling-Fong Li gave a colloquium at NTU on
the Higgs particle. At the end, I commented on the Higgs Mechanism
..., that Spontaneously Broken Gauge Theory (SBGT) of
SU(2)$\,\times\,$U(1) is already verified experimentally, and that
it does not require a physical Higgs boson. Then I said, roughly
``There should be enhanced chance for Japanese receiving the Prize
  this year, because former KEK Director General, Totsuka sensei,
  passed away in July.
  Second, BaBar shut down in April, marking a transition for the B Factory era.
  Therefore, as we await the LHC era to dawn, Cabibbo, Kobayashi and Maskawa
  have good chance this year.''

Turning around to my colleague, Yeong-Chuan Kao, a Nambu fan, I
said,
``Could it be Nambu and Goldstone\,?! Nambu certainly well
deserves, and he's really old !'' After all, his 65th birthday was
over 20 years ago. Kao then stated adamantly that Nambu will get
the prize.

On October 2nd, I gave a seminar at the IPMU of Tokyo University,
and stated after showing the recent CKM fit,
``Could this be the Year of CKM ?''
Hitoshi Murayama responded that ``the announcement should be
tomorrow.'' He got the date wrong, but it did prompt me to check
...

In the evening of October 7th, a little tired and bored, I
recalled the date and went on-line, and was exhilarated to witness
the announcement ... and the genius of the Nobel Committee. Though
``predicting'' all three Japanese names, I was certainly
astonished by the brutal cut-and-paste by the Committee.

So, 1/4 each of The Nobel Prize in Physics 2008 goes to Kobayashi
sensei and Maskawa sensei, ``\textit{for the discovery of the
origin of the broken symmetry which predicts the existence of at
least three families of quarks in nature.}'' The B factory workers
(this one included) feel as honored, by demonstrating the Standard
Model mechanism of CP Violation (CPV) is indeed as pointed out by
KM. Of course, by implication of the ordering, the more important
1/2 of The Nobel Prize in Physics 2008 goes to Nambu sensei,
``\textit{for the discovery of the mechanism of spontaneous broken
symmetry in subatomic physics.}'' (SSB) It is clear that SSB, the
``Higgs'' particle, the Origin of Mass, is the main goal at the
Large Hadron Collider.

I wonder whether the above can be called a ``premonition,'' but I
certainly did not expect to be here to deliver this ``lecture.''
But here goes.

\section{ Introduction: \ the Trouble of Mass(es) }

With Chadwick's discovery of the neutron, the near equality of
$m_n \cong m_p$ prompted Heisenberg to propose isospin symmetry,
that the two form an isodoublet, the nucleon $N$, and thus
introducing ``internal symmetries.'' From Yukawa's proposal to
Powell's discovery, as the dust settled, the pion also came in as
an isotriplet. But why, then, $m_\pi \ll m_N$?

It was Fermi and Yang who, in 1949, put forward the
suggestion~\cite{FermiYang} that maybe there is some compositeness
behind the scene. They pointed out that these particles were
assumed ``elementary,'' which essentially means ``structureless.''
But then they state that ``all such particles ... should be
elementary becomes less and less as their number increases.'' With
this, they suggest that the pion $\pi$ is a bound state of $N\bar
N$, but admitting immediately that $m_\pi \ll 2m_N$ is a real
difficulty.

By 1956, Fermi had died, while Yang was busy with parity
violation. It was Shoichi Sakata of Nagoya University who followed
up on the point. More particles have appeared since. Undaunted by
the mass issue, Sakata proposed~\cite{Sakata} that it is $N$ and
$\Lambda$ (or $p$, $n$ and $\Lambda$), an isodoublet and an
isosinglet, respectively, that are ``elementary,'' and all other
particles, whether the heavier hyperons, or the ``$\theta$'' or
``$\tau$'' mesons of the day, were composites \`a la the
Fermi--Yang suggestion.\footnote{
 Early indication of nucleon structure is the measurement of anomalous
 magnetic moment of proton by Otto Stern. At time of the Sakata paper,
 Hofstadter was uncovering nucleon structure at Stanford with
 $e$-beams.}

This was one step away from quarks. In the poster of ``The Jubilee
of the Sakata Model'' held at Nagoya University in October
2006~\cite{Sakata50}, it was shown, ingeniously, how $(\Lambda,\
p,\ n)$, turned upside down, became $(u,\ d,\ s)$, the quarks of
Gell-Mann. In the latter seminal paper~\cite{GMquark}, however,
Gell-Mann refers to ``the old Sakata model'' but without giving a
citation.

Still, where does the proton mass come from? [In fact, where does
``Mass'' come from? This is still the prevailing question of the
day.] And, can one understand $m_\pi \ll 2m_N$ with compositeness?
It was Nambu who guided the way, as documented~\cite{BrokenSymm}
in ``Broken Symmetry,'' which is also the namesake of the 2008
Nobel Prize in Physics.

\section{ Squalid State to the Rescue --- Nambu's Insight }

It was Gell-Mann who quipped the words ``Squalid State Physics,''
but it was Squalid State Physics that came to the rescue. Nambu,
who received a Sc.D. from Tokyo University in 1952, acknowledges
that~\cite{nambu-slides} ``my early exposure to condensed matter
physics has been quite beneficial to me.'' And Particle Physics
(and physics as a whole) is forever grateful.

Scientific American reran their ``Profile: Yoichiro Nambu'' of
1995, immediately after the Nobel announcement, with the
title~\cite{SciAm} ``Strings and gluons --- The seer, this year's
physics Nobel laureate, saw them all.'' Ed Witten is quoted as
saying ``\textit{People don't understand him, because he is so
farsighted.}'' And Murray Gell-Mann says ``\textit{Over the years,
you could rely on Yoichiro to provide deep and penetrating
insights on very many questions.}'' Sound like all praises.
However, in Nambu's own words, that he wrote decades later to
encourage a junior colleague:  ``\textit{Everyone seemed smarter
than I. I could not accomplish what I wanted to and had a nervous
breakdown.}'' He was referring to the Institute of Advanced Study,
the place he went after his doctorate. Even when he was already an
associate professor at Chicago, when he proposed a new particle in
1957, he was met with ridicule. Richard Feynman shouted
``\textit{In a pig's eye!}'' at the conference, recalls Laurie
Brown~\cite{LBrown}. The omega was discovered the next year.

But this was a momentous time, both at Chicago, and at Illinois.
``\textit{Give it another month, or a month and a half. Wait 'til
I get back and keep working. Maybe something'll happen.}'' With
these parting words to Bob Schrieffer, John Bardeen left for
Sweden in late November of 1956 to accept the Nobel Prize in
Physics, his first, for the discovery of the
transistor~\cite{BCSat50}. The rest is history. The BCS theory of
superconductivity was published in 1957~\cite{BCS}, a triumph of
Squalid State Physics!

In his autobiography notes in Broken Symmetry~\cite{BrokenSymm},
Nambu writes, ``\textit{One day before publication of the BCS
paper, Bob Schrieffer, still a student, came to Chicago to give a
seminar on the BCS theory in progress ... I was very much
disturbed by the fact that their wave function did not conserve
electron number. It did not make sense ... At the same time I was
impressed by their boldness and tried to understand the
problem.}'' Schrieffer soon joined the Chicago faculty, and Nambu
could discuss with him in depth. In July 1959, Nambu submitted the
paper~\cite{NambuSC} ``Quasi-particles and Gauge Invariance in the
Theory of Superconductivity.'' He obtained Ward identities
--- telltale signs of gauge invariance --- that, despite the
broken symmetry, rendered the Meissner effect calculation strictly
gauge invariant. He further observes that ``a pair of a particle
and antiparticle interacting with each other to form a bound state
with zero energy and momentum'' was crucial for understanding the
mechanism by which gauge invariance was restored. This is the
``Nambu-Goldstone boson'' of zero mass. Let us digress.

 \vskip0.25cm
\noindent\underline{Nambu-Goldstone Boson and Higgs Mechanism}
 \vskip0.25cm

Nambu later reflects~\cite{nambu-slides,Nambu07} on the mechanism
of BCS theory:
\begin{quote}
``The BCS theory assumed a condensate of charged pairs of
electrons or holes, hence the medium was not gauge invariant.
There were found intrinsically massless collective excitations of
pairs (Nambu-Goldstone modes) that restored broken symmetries, and
they turned into the plasmons by mixing with the Coulomb field.''
\end{quote}
He found that zero modes necessarily emerged because the broken
symmetry is continuous, the essence of the Goldstone theorem. Let
us not repeat the mathematics. It is well known that a ``Mexican
hat'' illustrates the point, so let me quote directly from Steven
Weinberg's talk ``From BCS to the LHC,'' delivered at the BCS@50
celebration~\cite{BCSat50}, and published in the CERN
Courier~\cite{WeinbergBCStoLHC}: ``Though the hat is invariant
under rotations about a vertical axis, a small ball will come to
rest off the axis of symmetry, somewhere on the brim of the hat,
but it can move freely with no restoring force around the brim.''
This is precisely the content of the Goldstone
Theorem~\cite{Goldstone}, made rigorous in a subsequent paper by
Goldstone, Salam and Weinberg. Though getting ahead of ourselves,
let me continue with Weinberg's words: ``Broken approximate
symmetry is illustrated by slightly tilting the hat; this produces
a small restoring force, analogous to the small mass of the
pion.''

The broken symmetry in superconductivity is not just continuous,
but it is the gauged symmetry of electromagnetism itself. Thus, as
we quoted from Nambu above, the ``Nambu-Goldstone modes ... turned
into the plasmons by mixing with the Coulomb field.'' This
explains the Meissner effect in a gauge invariant way (as insisted
by Nambu), which is nothing but the Higgs Mechanism.
The claim to the Higgs mechanism is of course a contentious one.
There is the argument based on BCS theory itself by Philip
Anderson~\cite{Anderson}, and three papers, by Fran\c{c}ois
Englert and Robert Brout, by Peter Higgs, and by Gerald Guralnik,
Carl Hagen and Tom Kibble~\cite{Higgs}.\footnote{
 We note in passing that Englert, Brout and Higgs received
 the EPS Prize in 1997, and the Wolf Prize in 2004.}

In reference to the Higgs mechanism, Nambu himself
uses~\cite{Nambu07} the term Ginzburg--Landau--Higgs ``effective''
field, citing the Ginzburg--Landau theory~\cite{GinzburgLandau},
the precursor to the BCS theory of superconductivity.
He further comments~\cite{nambu-slides}:
 ``\textit{In hindsight I regret that I should have explored in more
 detail the general mechanism of mass generation for the gauge field.
 But \underline{I thought the plasma} \underline{and the Meissner effect had
 already established it}. I also should have paid more attention to
 the Ginzburg-Landau theory which was a forerunner of the present
 Higgs description.}''\footnote{
 The underlining is mine.}
On other examples of BCS type of SSB, such as $^3$He superfluidity
and nucleon pairing in nuclei, he offers his opinion on fermion
mass generation in the Standard Model~\cite{nambu-slides}:
 ``\textit{my biased opinion, there being other interpretations
 as to the nature of the Higgs field}.''
Let me now return to Nambu's prize.

 \vskip0.25cm
\noindent{\underline{The Penetrating Analogy}: \
 $m_N$ as ``BCS Energy Gap'' and $\pi$ as ``collective excitation''}
 \vskip0.25cm

Nambu followed a keen interest in understanding gauge invariance
in the presence of SSB. He was certainly more a particle and field
theorist, than a condensed matter physicist. With the proposal of
$V-A$ theory of weak interactions and CVC in 1958, many were
interested in a possibly conserved axial vector current. In the
paper~\cite{NambuA} ``Axial Vector Current Conservation in Weak
Interactions,'' Nambu constructed the axial vector current, and
having derived the Goldberger--Treiman relation in a satisfactory
way, he made the penetrating analogy from his understanding of BCS
theory:
gauge invariance $\leftrightarrow$ $\gamma_5$ invariance; energy
gap $\leftrightarrow$ baryon mass; and collective excitations
$\leftrightarrow$ mesons.
He went on to conclude that ``It is interesting that pseudoscalar
mesons automatically emerge in this theory as bound states of
baryon pairs. The nonzero meson masses and baryon mass splitting
would imply that the $\gamma_5$ invariance of the bare baryon
field is not rigorous, possibly because of a small bare mass of
the order of the pion mass.'' Although the last statement was not
entirely correct, the insight was clearly profound: what was
behind the Fermi--Yang--Sakata model was becoming clearer!
Yukawa's pion has become the Nambu--Goldstone boson of spontaneous
Chiral Symmetry Breaking ($\chi$SB), and light had been shed on
why $m_\pi \ll 2m_N$. The dominant part of our body mass, or that
of the Earth or the Sun or the visible Universe, arose through
$\chi$SB.

%
\begin{table}[htb]
\centering
 \caption{
  Nambu's analogy between the BCS theory of superconductivity,
  and spontaneously broken chiral symmetry of the strong interactions
  (adapted from Ref.~\cite{phyadv}).}
 \label{tab:NambuSSB}       
 \vskip0.15cm
\centering
\begin{tabular}{|c|c|c|c|}
\hline 
 & \ \ Superconductivity\ \ & \ \ Strong Interactions\ \ & \ \ realization \ \  \\
\hline 
 \ \ elementary particle\ \  & electrons
                          & \ \ hypothetical fermions \ \ & \ \ $u$, $d$ quarks \ \ \\
\hline
 interaction\ \           & phonon exchange & unknown & \ QCD \ \\
\hline
                          & energy gap & $m_N$ &  \\
\hline
                          & \ \ collective excitations \ \
                          & \ \ $\pi$ (meson boundstate) \ \ &  \\
\hline
                          & electric charge
                          & \ \ chirality \ \ &  \\
\hline
 broken symmetry  & gauge invariance
                          & \ \ chiral ($\gamma_5$) invariance \ \ &  \\
\hline
\end{tabular}
\end{table}

Table~\ref{tab:NambuSSB} illustrates the full power of Nambu's
analogy. Considering that the ``hypothetical fermions'' were put
forth 4 years before the quark model~\cite{GMquark,Zweig}, and the
fact that it took another decade (after the emergence of QCD that
took the place of the ``unknown'' interaction) for the latter to
be accepted, Nambu's insight by analogy is truly penetrating. He
saw through the veil of what baryons and mesons really are, even
before the rather difficult dynamics was understood. We should
remember that color SU(3) was also proposed by
him~\cite{HanNambu}, on the grounds of statistics. Here he
followed Gell-Mann's quark model paper, but unfortunately did not
follow the simplest possibility of fractionally charged quarks.
With the latter, we have the familiar picture that the proton
(neutron) has $uud$ ($udd$) constituents, whereas $\pi^+$ is
$u\bar d$.

Although the true dynamics was not yet known, Nambu went further
to construct dynamical chiral symmetry breaking, eventually
resulting in the Nambu--Jona-Lasinio (NJL) model~\cite{NJL}, and
the foundations of chiral perturbation theory ($\chi$PT). For
clarity, perhaps we need to backtrack to what Nambu saw in the
nucleon with $\chi$SB vs. BCS theory, and we would need a few
formulas.

The quasi-particles near the Fermi surface of a superconductor is
a coherent mixture of electron and hole states with momentum $p$
and spin $+$,

\begin{eqnarray}
 E\,\psi_{p,+} &=& \phantom{-}\epsilon_p\,\psi_{p,+} \ \;\,+
 \phi\,\psi^\dagger_{-p,-}, \nonumber\\
 E\,\psi^\dagger_{-p,-} &=& -\epsilon_p\,\psi^\dagger_{-p,-} +
 \phi\,\psi_{p,+}, 
  \\
 E &=& \pm\sqrt{\epsilon_p^2 + \phi^2}, \nonumber
\end{eqnarray}
where $\epsilon_p$ is the kinetic energy from the Fermi surface,
and $\phi$ is the gap parameter~\cite{nambu-slides,NJL}.
Nambu saw an analogy with the (relativistic) Dirac equation, which
reads, in the Weyl representation,
\begin{eqnarray}
 E\,\psi_1 &=& \phantom{-}\sigma\cdot p\,\psi_1 +
 m\,\psi_2, \nonumber\\
 E\,\psi_2 &=& -\sigma\cdot p\,\psi_2 +
 m\,\psi_1, 
  \\
 E &=& \pm\sqrt{p^2 + m^2}, \nonumber
\end{eqnarray}
where $\psi_{1,2}$ are eigenstates of the chirality operator
$\gamma_5$. The Dirac mass corresponds to the gap parameter
$\phi$, which touches base with our former discussion. But from
here, NJL states~\cite{NJL} ``As the energy gap $\phi$ in a
superconductor is created by the interaction, let us assume that
the mass of a Dirac particle is also due to some interaction
between massless bare fermions.''\footnote{
 The relativistic cutoff would take the place of the
 (nonrelativistic) Fermi surface.}
Going further, ``It is perhaps not a coincidence that there exists
such an entity in the form of the pion. For this reason, we would
like to regard our theory as dealing with nucleons and mesons. The
implication would be that {\it the nucleon mass is a manifestation
of some unknown primary interaction between originally massless
fermions, the same interaction also being responsible for the
binding of nucleon pairs into pions}.'' (italics are mine) This is
the starting point not only in constructing the NJL model, but
$\chi$PT, and effective field theories in general. Note that
$\chi$PT works at long distance when QCD doesn't, and even lattice
QCD calculations call on $\chi$SB to interface with data.

We see the power of one person's insight through analogy.

\section{ Spontaneous Symmetry Breaking and SU(3)$\,\times\,$SU(2)$\,\times\,$U(1) }

SSB is clearly an integral part of the Standard Model, both in
dynamical $\chi$SB, and the SSB of electroweak gauge interactions.

With the ``unknown'' interaction of spontaneous/dynamical $\chi$SB
established as the non-Abelian gauged dynamics of color SU(3), we
recall the breakthrough of asymptotic freedom that preceded it,
and the correct selection of the gauge group by Gross and Wilczek
(and Fritzsch and Gell-Mann), even though Han and Nambu had
proposed it much earlier. So, the global SU$_{\normalsize\rm
L}$(3)$\,\times\,$SU$_{\normalsize\rm R}$(3) chiral symmetry is
dynamically broken down by QCD, as well as explicit quark masses,
to approximate SU$_{\normalsize\rm V}$(3). But, just as in the NJL
model, even when QCD is not applicable in the strong coupling
regime (low energy), $\chi$PT is applicable. This is the precursor
to effective field theories.

We are familiar with Weinberg's usage of the Higgs mechanism to
break SU(2)$\,\times\,$U(1), and his conjecture that non-Abelian
gauge theory is renormalizable, and that reordering of
perturbation after SSB would not change the
renormalizability~\cite{Weinberg}. These were of course later
proven by 't~Hooft and Veltman.
What we wish to stress is that, while a single complex Higgs
doublet field developing a vacuum expectation value (v.e.v.) is
the simplest realization of SU(2)$\,\times\,$U(1) breaking, the
Higgs field does not have to be elementary. One needs only an
effective Ginzburg--Landau--Higgs field~\cite{nambu-slides}; SSB
is already an experimental fact, and regardless of the true origin
of the SSB, the longitudinal $W$ and $Z$ are the would-be
Nambu--Goldstone bosons!

\section{ Gell-Mann--L\'evy--Cabibbo and GIM: \ from 3 to 4 Quarks }

The proliferation of strange hadrons caused Sakata to propose his
substructure model, their weak decay also pointed to some
universal behavior.

In his seminal paper ``Unitary Symmetry and Leptonic Decays'' of
1963, Nikola Cabibbo made a comprehensive study of strangeness
conserving ($\Delta S = 0$) and changing ($\Delta S = 1$)
decays~\cite{Cabibbo}. Writing before quarks, he used SU(3)
currents and $V-A$ theory to compare e.g. $\pi^+$ vs. $K^+$ decay
to $\mu^+\nu_\mu$. He proposed a ``unit length'' of weak
interaction strength, i.e. a $\cos\theta$, $\sin\theta$ factor for
the $\Delta S = 0$, 1 currents, and found that a near universal
value of $\theta \simeq 0.26$ could account for available data. He
noted that, though overshooting a bit, the $\cos\theta$ factor to
$^{14}$O 
beta decay was in the right direction. Paraphrasing the Nobel
Committee~\cite{phyadv}, ``the Cabibbo Theory, with the Cabibbo
angle,'' $\theta_C \cong 13^\circ$, ``quickly became a standard
framework for the weak interactions. It turned out to be universal
and an ever-increasing multitude of data could be fitted into it.
It has been a cornerstone of weak interactions.'' In quark
language, the $d$ quark in weak interactions is a mixture,
\begin{equation}
 d' = d\, \cos\theta_C + s\, \sin\theta_C,
\end{equation}
of the $d$ and $s$ mass eigenstates.

Unfortunate perhaps for Cabibbo, the unitary transform of $\Delta
S = 0$ and 1 currents had been consider 3 years earlier by
Gell-Mann and Maurice L\'evy~\cite{GML60} in their work ``The
Axial Vector Current in Beta Decay,'' as Cabibbo acknowledges in
his footnote~4. Gell-Mann and L\'evy noted that $^{14}$O data
implied $G/G_\mu \sim 0.97$. In a ``{\it Note added in proof,}''
they suggested to
\begin{quote}
``... consider the vector current for $\Delta S = 0$ and $\Delta S
= 1$ together to be something like:
\[
 G\,V_\alpha + G\,V_\alpha^{(\Delta S = 1)}
  = G_\mu\, \bar
  p\gamma_\alpha(n+\varepsilon\,\Lambda)(1+\varepsilon^2)^{-\frac{1}{2}}
  + \cdots, 
\]
and likewise for the axial vector current. If
$(1+\varepsilon^2)^{-\frac{1}{2}} = 0.97$, then $\varepsilon^2 =
.06$, which is of the right order of magnitude for explaining the
low rate of $\beta$ decay of the $\Lambda$ particle.''
\end{quote}
to maintain universality. Although they did not press further,
this is the same program as that pursued by Cabibbo (who had more
data available), and may be why the Nobel Committee uses the
wording ``Gell-Mann--L\'evy--Cabibbo'' theory~\cite{phyadv},
something that is not in common use.

In 1964 Gell-Mann put forth the quark model~\cite{GMquark}, where,
dispensing with the need of a ``basic neutral baryon singlet,'' he
discusses the simpler (and calling it more elegant) non-integer
charged ``quarks'' (here, Gell-Mann is moving beyond the Sakata
model). With the $u$, $d$ and $s$ quarks, Gell-Mann puts the
electromagnetic current in the modern form, and uses the
\textit{quark mixture} of Eq.~(4.1) for the first time. In so
doing, Gell-Mann refers both to his previous work with L\'evy, as
well as stating ``We thus obtain all the features of Cabibbo's
picture.''

Besides the experimental discovery~\cite{CPV64} of CPV in
$K_L^0\to \pi\pi$, on the path to Kobayashi and Maskawa, there is
one more important step, which is the mechanism proposed by
Sheldon Glashow, Jean Iliopoulos and Luciano Maiani
(GIM)~\cite{GIM} in 1970, in the paper titled ``Weak Interactions
with Lepton--Hadron Symmetry.'' If the $d'$ of Eq.~(4.1) enters
the charged current with the $u$ quark, then in the
Glashow--Weinberg--Salam theory, 
the $Z^0$ boson would couple to a $\bar d'd'$ pair, and $K^0$
(composed of $\bar sd$) should decay to $\mu^+\mu^-$, which was
not observed. GIM pointed out that, if one adds the orthogonal
combination\footnote{
 GIM uses the Sakata model notation of ${\cal P,\ N},\ \lambda$
 for the SU(3) triplet of quarks.}
of $s' = -d\, \sin\theta_C + s\, \cos\theta_C$, then the $K^0(\bar
sd) \to \mu^+\mu^-$ amplitude cancels away. This works also at the
loop level (violated only by difference in masses), guaranteeing
that flavor changing neutral currents (FCNC) are suppressed, which
is consistent with experiment! This usage of unitarity of quark
mixing matrix is the foundation of flavor loop calculations.

The $Z^0$ coupling to $\bar s's'$ pair implies the existence of a
new charge $+2/3$ quark. This ``charm'' quark, $c$ in modern
notation (${\cal P}'$ as used by GIM), renders the ``Cabibbo
rotation'' complete.

\section{ Insight of Kobayashi--Maskawa: \ 6 Quarks for CPV }

The experimental discovery~\cite{CPV64} of CP violation in neutral
kaon system in 1964 came as a surprise. Before long, however,
Andrei Sakharov pointed out~\cite{Sakharov} that CPV is needed to
explain the absence of antimatter in our Universe, so
understanding CPV gained \textit{universal} importance.

The GIM paper did mention~\cite{GIM} ``... suitable redefinitions
of the relative phases of the quarks may be performed in order to
make $U$ real and orthogonal ...,'' and in this way they arrived
at the $2\times 2$ rotation matrix. But they were not concerned
with CP violation.

The pursuit of CP violation was taken up by Makoto Kobayashi and
Toshihide Maskawa (KM). Both received their Ph.D. from Nagoya
University (hence immersed in the Sakata school), and both went to
Kyoto University after graduation. Barely 5 months after Kobayashi
had arrived, the two submitted the paper ``CP Violation in the
Renormalizable Theory of Weak Interaction''~\cite{KM} on September
1st, 1972 to Progress of Theoretical Physics, the Japanese
journal. Kobayashi was just 28 years old, and Maskawa 32. Although
the title states ``renormalizable,'' they refer only to Weinberg's
paper~\cite{Weinberg}.
Nor do they refer to GIM,\footnote{
 Knowledge of GIM was acknowledged in the Nobel Lecture of
 Maskawa~\cite{maskawa-slides}.}
but they also use Sakata's notation of $p$, $n$, $\lambda$, plus
$\zeta$ for the 4th, charge $+2/3$ quark (the charge was actually
kept free in the work). At Nagoya, the school of Sakata, people
were independently thinking of the existence of a 4th quark after
$\nu_\mu$ was demonstrated to be different from $\nu_e$. So it
came natural to the two Nagoya graduates.

KM quickly dispensed with the ``quartet'' model, that ``With an
appropriate phase convention of the quartet field we can take $U$
as ...'' the familiar $2\times 2$ form. Targeting CPV, they
conclude that ``no realistic models of $CP$-violation exist in the
quartet scheme without introducing any other new fields.'' So they
consider adding new fields, such as extra scalar
doublets.\footnote{
 The case of extra right-handed quark doublets were already
 considered by Rabindra Mohapatra~\cite{Rabi}.
 }
As for ``a 6-plet model, another interesting model of
$CP$-violation,'' they note that ``in this case we cannot absorb
all phases of matrix elements into the phase
convention...,''\footnote{
 I learned from the Nobel Committee~\cite{phyadv}, that our usual
 exercise (in our Particle Physics course) of phase freedom of
 quark fields, the $n^2 - (2n-1) = (n-1)^2$ degrees of freedom in
 the $n\times n$ $V_{\rm CKM}$ matrix, is something called the
 Iwasawa decomposition~\cite{Iwasawa} of the unitary matrix $U$
 that connects two $n$-component vector fields.}
and they give their parametrization of the $3\times 3$ unitary
matrix. The 3 ``generation'' Standard Model was born ---
\textit{motivated by CPV}.

I bless the young postdoc(s) reading this to make similar impact
at a tender age.

\section{ Richness of KM: \ from tau to top; FPCP }

People often think Nambu as a cut above the other two recipients,
and I concur. Nambu is a cut above most of us, maybe even a good
fraction of the Nobel recipients. But the measure of success of a
model is its predictive impact! And here, the KM model --- the
flavor (and CPV) part of the Standard Model --- is indeed very
successful. If Cabibbo captured half of the $2\times 2$ quark
mixing matrix, and he could account for practically all (tree
level) strangeness decay and production with one parameter, then
KM predicted the existence of two additional mixing angles, plus
the crucial CPV phase to account for that enigmatic phenomena.
There is the underlying prediction of the existence of the third
generation $\tau$ (and $\nu_\tau$) lepton and $b$ and $t$ quarks.
Though the values are not predicted, masses and mixings account
for the major number of --- dynamical --- parameters in the
Standard Model, which calls out for an explanation. This
conference --- Flavor Physics and CP Violation --- is just shaping
the foundation for the next step, i.e. understanding it all. The
discovery of the third generation fermions, measuring their
properties, confirming that it is indeed the source of CPV, is the
enterprise that honors the insight of two young men. It is mostly
the outcome of the B factories, as epitomized in the ``CKM
unitarity fit'' (salutations to Cabibbo again), that lead to KM's
Prize. This is our payback to these two gentlemen. The fact they
did it from a ``remote'' part of the academic world just adds to
the allure. The observation {\it did not} come from the big wigs
of the West (or Russia?). This brings me, an East Asian,
satisfaction.

There is unfinished business, however.

Towards the end of Kobayashi's Nobel
Lecture~\cite{kobayashi-slides}, he comments on the state of
affairs of CP violation studies. Though ``B-factory results show
that quark mixing is the dominant source of CP violation,'' they
``allow room for additional source from new physics.'' This calls
for further studies. What is even more enticing is that ``Matter
dominance of the Universe seems requiring new source of CP
violation.'' Let us elucidate this a little.

What the B factories have checked in detail is that
\begin{equation}
 V_{ud}V_{ub}^* + V_{cd}V_{cb}^* + V_{td}V_{tb}^* = 0
\end{equation}
forms a \textit{nontrivial} triangle, and a nontrivial CPV phase
in particular. However, at the more refined level, the Iwasawa
decomposition for the $3\times 3$ case demands nondegeneracy of
the bi-vector space; in physical terms this means all like-charge
quark pairs ought to be nondegenerate. Otherwise, one is
effectively back to the 2 generation case, and the CPV phase can
be removed. Cecilia Jarlskog captured the essence of this through
her proposed invariant measure of CPV~\cite{Jarlskog},
\begin{equation}
J = (m_t^2 - m_u^2)(m_t^2 -  m_c^2)(m_c^2 - m_u^2)
    (m_b^2 - m_d^2)(m_b^2 -  m_s^2)(m_s^2 - m_d^2)\, A,
\end{equation}
which can be deduced from
${\rm Im}\,\det\left[ m_u m_u^\dagger,\ m_d m_d^\dagger
               \right]$, where
$A$ is twice the area of the triangle of Eq.~(6.1). Indeed, one
sees that $J$ vanishes with $A$, or when any pair of like-charge
quarks are degenerate.
Kobayashi's words that ``Matter dominance of the Universe seems
requiring new source of CP violation'' refers to the despairing
situation that $J$, normalized to the weak scale, seems to fall
short of what is needed for the Sakharov conditions by at least a
factor of $10^{-10}$.

\section{ Epilogue: \ KM--N Redux?  A Perspective }

Nambu had a critical opinion on the nature of the Higgs field and
fermion mass generation, while Kobayashi noted that his CPV phase
with Maskawa, the one confirmed by the B factories, falls short of
what is needed for matter dominance of the Universe. I then ask:
Can there be a redux of Kobayashi--Maskawa--Nambu? In the
following, I turn to offer a (personal) perspective, with the
caution that --- \textit{I'm a born-again 4th generationist!} Stop
here if you wish.

At this conference, considerable effort is spent on the
theoretical understanding of charmless hadronic $B$ decays. The
prevailing wind since a few years is to brush aside the
staggering, \textit{unpredicted} direct CPV difference of
$\Delta {A}_{K\pi} \equiv
 {\cal A}_{B^+\to K^+\pi^0} - {\cal A}_{B^0\to K^+\pi^-}
 > -{\cal A}_{B^0\to K^+\pi^-} \sim 10\%$,
as due to a large, hadronically enhanced color-suppressed
amplitude $C$.\footnote{
 To me, this mouthful, as well as the oxymoronic
 ``enhanced color-suppressed ...,'' are clear symptoms.}
But, shocked by the emergence of $\Delta {A}_{K\pi}$ in 2004, I
demonstrated~\cite{HNS}, together with Makiko Nagashima and Andrea
Soddu, that it could be due to the 4th generation: the
nondecoupled $m_{t'}^2$ dependence of the $bsZ$ penguin loop is
tailor-made, while $V_{t's}^*V_{t'b}$ brings in a new CPV phase.
Stressing that the $bsZ$ penguin and the $b\bar s \leftrightarrow
s\bar b$ box diagrams are closely linked, we made a prediction of
$\sin2\Phi_{B_s}$ (defined analogously to $\sin2\beta/\phi_1$, but
for $B_s\to J/\psi \phi$) on the basis of a finite $t'$ effect in
$\Delta {A}_{K\pi}$. The prediction was improved~\cite{HNS07} to
$\sin2\Phi_{B_s} \simeq -0.5$ to $-0.7$ with a somewhat different
argument, after $\Delta m_{B_s}$ was measured by the CDF
experiment in 2006.

Encouraging results followed at the Tevatron. I am grateful to CDF
for the wording~\cite{CDFpublic} ``... some source of new physics
contributing to the electroweak penguin which governs the $b \to
s$ transition. ... George Hou predicted the presence of a $t'$
quark ... to explain the Belle result and predicted \textit{a
priori} the observation of a large \textit{CP}-violating phase in
$B_s^0 \to J/\psi\,\phi$ decays [7,\ 8].'' This refers to our
prediction of $\sin2\Phi_{B_s}$. Although the (combination of)
errors are still under discussion, it is astounding that the world
central value of 2008 is about $-0.6$, with significance above
2$\sigma$, and deviating that much from the Standard Model
(predicted from replacing $d$ by $s$ in the 3 generation relation
of Eq.~(6.1)). This is astounding, as theorists only look at
central values.

A little more than a year ago, I made an unsettling observation. I
learned about Jarlskog's invariant as a young postdoc, but only
through the writing of the Belle \textit{Nature}
paper~\cite{BelleNature} on $\Delta {A}_{K\pi}$ did I learn more
clearly the simple dimensional analysis argument that, when scaled
by the electroweak phase transition temperature of $T \sim 100$
GeV (worse if v.e.v. is used), then $J$ falls short of $n_{\cal
B}/n_\gamma \sim 0.5 \times 10^{-9}$ by a factor of $10^{-10}$. It
occurred to me in late summer 2007, which I put forth in
arXiv:0803.1234, that if there is a 4th generation, and if one
shifts\footnote{
 This is valid in the $d$-$s$ degeneracy limit, which
 is justified in our world.}
from $123$ to $234$ in Eq.~(6.2), then ``$J_{234}$'' gains a
factor of $\sim 10^{15}$~\cite{CPVBAU}. The gain is mostly in
large Yukawa couplings. \textit{It seems that one would have
enough CPV for matter dominance of the Universe}, and Nature would
likely use this. Who can argue with the ``.1234'' number returned
by arXiv on March 8, 2008?

Another thing I enjoyed learning in the same time frame, prior to
Nambu getting his prize, is that heavy $\bar QQ$ can condense
through (the origins of) large Yukawa coupling. Concurrent with
Nambu's ``biased opinion,'' the Higgs boson would then become
composite. Could SSB of electroweak gauge symmetry be due to
condensation of $b'$/$t'$ quarks near or above the unitarity bound
of 500--600 GeV? Bob Holdom has explored~\cite{Holdom} this theme
in the past few years ($\sim$ the Bardeen--Hill--Lindner work 20
years ago), which is in fact nothing but the NJL model!!
Interestingly, there is an AdS/CFT correspondence to this strong
coupling regime, called~\cite{Burdman} the ``holographic 4th
generation.'' I was therefore more than pleased with Nambu
receiving the 2008 Nobel Prize.

So, are we in for a KM--N redux? The $10^{15}$ gain in CPV may be
enough as the source of CPV for the matter dominance of the
Universe (remember, the fermion masses are really Yukawa
couplings, hence the dynamical effect may work even above the
electroweak phase transition scale). This alone to me is reason
enough that ``\textit{Maybe there is a 4th Generation!}\,'' It is
tantalizing that the Tevatron could bring indirect evidence for
this through $\sin2\Phi_{B_s}$ by 2010-2011; it could be easily
pursued at LHCb, once data arrives. More important, direct search
at the LHC by the CMS and ATLAS experiments could discover, or
rule out, the existence of a 4th generation in 3-5 years
time~\cite{4-4th}. If found, we would have brought heaven on
earth, and with luck, the riddle of electroweak symmetry breaking
could be unveiled by the heaviness of the new chiral quarks.

And that would be the ultimate tribute to the insights of
Kobayashi, Maskawa, and Nambu.

\end{document}